\documentclass[iop]{emulateapj}

\usepackage{graphicx}
\usepackage{amssymb}
\usepackage{amsmath}
\usepackage{epstopdf}

\def \arcsec {\ensuremath{^{\prime\prime}}}
\def \rxj {RX\,J0520.5$-$6932}
\def \nustar {\emph{NuSTAR}}
\def \swift {\emph{Swift}}
\def \xmm {\emph{XMM-Newton}}
\def \fermi {\emph{Fermi}}
\def \msun {\ensuremath{M_\odot}}

\shorttitle{\nustar~ Observations of \rxj}
\shortauthors{Tendulkar, S.~P. et al.}

\begin{document}

\setlength{\fboxsep}{0pt}%
\setlength{\fboxrule}{0.5pt}%

\title{\emph{NuSTAR} Discovery of a Cyclotron Line in the Be/X-ray Binary RX\,J0520.5$-$6932 During Outburst}

\author{Shriharsh P. Tendulkar\altaffilmark{1}, Felix F\"urst\altaffilmark{1}, Katja Pottschmidt\altaffilmark{2,3}, Matteo Bachetti\altaffilmark{4,5}, Varun B. Bhalerao\altaffilmark{6}, Steven E. Boggs\altaffilmark{7}, Finn E. Christensen\altaffilmark{8}, William W. Craig\altaffilmark{9}, Charles A. Hailey\altaffilmark{9}, Fiona A. Harrison\altaffilmark{1}, Daniel Stern\altaffilmark{10}, John A. Tomsick\altaffilmark{7}, Dominic J. Walton\altaffilmark{1}, William Zhang\altaffilmark{3}}

\email{spt@astro.caltech.edu}

\altaffiltext{1}{Space Radiation Laboratory, California Institute of Technology, 1200 E California Blvd, MC 249-17, Pasadena, CA 91125, USA}
\altaffiltext{2}{Center for Research and Exploration in Space Science and Technology, University of Maryland, Baltimore County, Baltimore, MD 21250, USA}
\altaffiltext{3}{NASA Goddard Space Flight Center, Astrophysics Science Division, Code 661, Greenbelt, MD 20771, USA}
\altaffiltext{4}{Universit\'e de Toulouse, UPS-OMP, IRAP, F-31400 Toulouse, France}
\altaffiltext{5}{CNRS, Institut de Recherche en Astrophysique et Plan\'etologie, F-31028 Toulouse, Cedex 4, France}
\altaffiltext{6}{Inter-University Centre for Astronomy and Astrophysics, Post Bag 4, Ganeshkhind, Pune University Campus, Pune 411 007, India}
\altaffiltext{7}{Space Sciences Laboratory, University of California, Berkeley, CA 94720, USA}
\altaffiltext{8}{DTU Space, National Space Institute, Technical University of Denmark, Elektrovej 327, DK-2800 Lyngby, Denmark}
\altaffiltext{9}{Columbia Astrophysics Laboratory, Columbia University, New York, NY 10027, USA}
\altaffiltext{10}{Jet Propulsion Laboratory, California Institute of Technology, Pasadena, CA 91109, USA}


\begin{abstract}
We present spectral and timing analysis of \nustar\ observations of \rxj\ in the 3--79\,keV band collected during its outburst in January 2014. The target was observed on two epochs and we report the detection of a cyclotron resonant scattering feature with  central energies of $E_\mathrm{CRSF} = 31.3_{-0.7}^{+0.8}$\,keV and $31.5_{-0.6}^{+0.7}$\,keV during the two observations, respectively, corresponding to a magnetic field of $B \approx 2 \times10^{12}$\,G. The 3--79\,keV luminosity of the system during the two epochs assuming a nominal distance of 50\,kpc was $3.667\pm0.007\times 10^{38}\,\mathrm{erg\,s^{-1}}$ and $3.983\pm0.007\times10^{38}\,\mathrm{erg\,s^{-1}}$. Both values are much higher than the critical luminosity of $\approx1.5\times10^{37}\,\mathrm{erg\,s^{-1}}$ above which a radiation dominated shock front may be expected. This adds a new object to the sparse set of three systems that have a cyclotron line observed at luminosities in excess of $10^{38}\,\mathrm{erg\,s^{-1}}$. A broad ($\sigma\approx0.45$\,keV) Fe emission line is observed in the spectrum at a central energy of $6.58_{-0.05}^{+0.05}$\,keV in both epochs. The pulse profile of the pulsar was observed to be highly asymmetric with a sharply rising and slowly falling profile of the primary peak. We also observed  minor variations in the cyclotron line energy and width as a function of the rotation phase.
\end{abstract}

\maketitle

\section{Introduction}
\rxj\ is a Be/X-ray binary discovered in a \textit{ROSAT} survey of the Large Magellanic Cloud \citep{schmidtke1994}. Optical spectroscopy suggested a companion star of O8Ve spectral type with radial velocity measurements consistent with LMC membership. An analysis of photometric data from the Optical Gravitational Lensing Experiment \citep[OGLE; ][]{udalski1992} revealed a 24.4\,d periodicity in the $R$-band luminosity. Later spectro-photometric data improved the spectral identification to O9Ve \citep{coe2001} corresponding to a companion mass between 17--23\,\msun.

\begin{figure}
\center
\includegraphics[clip=true,trim=0.2in 0in 0.2in 0in,width=0.48\textwidth]{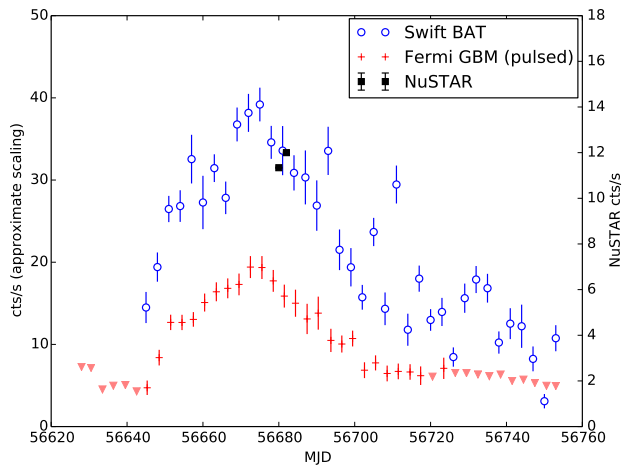}
\caption{The \swift /BAT (15-50\,keV) lightcurve (blue circles) and \emph{Fermi}/GBM pulsed flux measurement (red `+'s) and non-detections (red inverted triangles). The two count rates have been approximately scaled using the BAT and GBM nominal collecting areas and energy ranges for plotting convenience. The black squares show the 3--79\,keV \nustar\ count rates measured in Observation I and II. The count rates have been corrected by the \nustar\ pipeline for detector losses. The slight increase in the \nustar\ count rates cannot be confirmed from by averaging the \swift-BAT lightcurve over the \nustar\ observations due to the lack of sufficient statistics.}
\label{fig:lightcurve}
\end{figure}

In January 2013, a \swift-XRT survey of the LMC revealed \rxj\ to have entered an X-ray outburst, the first since its discovery, with a 0.2--12\,keV flux $\approx$25 times higher than the previous measurement \citep{vasilopoulos2013ATela,vasilopoulos2014}. Subsequent \xmm\ and \swift/XRT observations revealed a spin period of 8.034(5)\,s \citep{vasilopoulos2013ATelb}. In early January 2014, the source reached a sustained 0.3--10\,keV X-ray luminosity of $L_X \approx1.91\times10^{38}\,\mathrm{erg\,s^{-1}}$, approximately equal to the Eddington luminosity of an accreting neutron star \citep{vasilopoulos2014ATel}. Pulsations were detected in the 12--25\,keV band by the \fermi/GBM NaI detectors, and were monitored \citep{finger2009} over a period between 2013, December 18 and 2014, March 6  (Figure~\ref{fig:lightcurve}). \citet{kuehnel2014ATel} used the Doppler variation in the \fermi/GBM measurement of the pulsar spin period to fit the orbit, finding an orbital period, $P_\mathrm{orb} = 23.93(7)$\,d and a semi-major axis, $a\sin i = 107.6(8)$\,lt-sec. Assuming a stellar companion mass of 17--23\,\msun (appropriate for the spectral type), this corresponds to an orbital inclination of 27--31$^{\circ}$.

\emph{Nuclear Spectroscopic Telescope ARray} \citep[\nustar;][]{harrison2013} observations of \rxj\ were performed with a goal of obtaining high-resolution hard X-ray spectrum for this source in a rare near-Eddington accretion state. \nustar's sensitivity and spectral resolution make it feasible to observe and resolve possible cyclotron absorption features at a 3--79\,keV flux level of $\sim10^{-9}\,\mathrm{erg\,cm^{-2}\,s^{-1}}$. In this paper, we describe the detection of a cyclotron absorption line in phase-resolved observations during the two epochs. The paper is organized as follows: in Section~\ref{sec:obs} we describe the \nustar\ observations and data analysis procedure, in Section~\ref{sec:results} we describe the spectral modeling and timing analysis and in Section~\ref{sec:discussion} we discuss the implications of our measurements.

\subsection{Cyclotron Resonant Scattering Features}
Cyclotron resonant scattering features (CRSFs, also known as cyclotron absorption lines) can be formed in accretion columns of highly magnetized neutron stars and are most often observed in High Mass X-ray Binaries (HMXBs)\footnote{Only two Low Mass X-ray Binaries (LMXBs) are known to show CRSFs: 4U\,1626$-$67  \citep{orlandini1998}
 and Her X-1 \citep{trumper1978}
.} where active accretion onto polar regions of neutron stars is occuring \citep[see][for a review]{caballero2012}. CRSFs offer the only direct measurement of magnetic fields near the neutron star surface. These features, usually observed in absorption against a continuum spectrum, are created due to the increased scattering cross-section of plasma particles to photons of energy equal to the gap between Landau levels. The energy gaps between the Landau levels, for electrons, are given by the so-called 12-$B$-12 rule: $E_\mathrm{CRSF} \approx 11.57\times B_{12}(1+z)^{-1}\,\mathrm{keV}$, where $E_\mathrm{CRSF}$ is the energy gap, $B_{12}$ is the magnetic field in units of $10^{12}$\,G and $z$ is the gravitational redshift from the neutron star \citep[see ][and references therein]{schonherr2007}. 

Physically, the continuum emission is thought to arise from a shock boundary where the accreting material, funneled towards the poles by the magnetic field, slows from supersonic to subsonic flow. \citet{becker2012} suggested that at low accretion rates the shock is dominated by the Coulomb interactions in the plasma, and hence for increasing mass accretion rate (i.e. increasing luminosity) the shock location moves closer to the surface and encounters higher magnetic fields. Above a critical accretion rate (with corresponding luminosity $L_\mathrm{crit}$), the shock becomes dominated by radiation travelling upwards through the accretion column, and the shock location height increases as the accretion rate increases, causing an anti-correlation between $B_{12}$ and $L_X$. Recent work by \citet{nishimura2014} suggests that variation in the accretion column radius and the direction of photon propagation may also play a significant role in explaining the observed anti-correlation. The expected correlation (at low luminosities) and anti-correlation (at high luminosities) is qualitatively corroborated by current observations \citep[see Figure 6 in][]{furst2014_velax1}, although quantitatively some variations remain unexplained. Only a few sources exist with CRSF observations at $L_X>L_\mathrm{crit}$: V0332+53 \citep{coburn2005ATel,kreykenbohm2005,pottschmidt2005,mowlavi2006,tsygankov2006}, KS\,1947+300 \citep{furst2014_ks1947} and 4U\,0115+63 \citep[e.g.]{wheaton1979,muller2013}. The detection presented in this paper adds a new data point to this sparse set. 

 The continuum emission in X-ray pulsars is usually well-fit by a powerlaw spectrum with an exponential cutoff corresponding to the highest electron energies. The photons arising from the shock travel to the observer through the surrounding magnetized plasma, which imprints the characteristic cyclotron resonant scattering feature as a broad absorption trough. As the neutron star rotates, the observer's line of sight passes through regions with varying magnetic field strength, varying the energy, width and depth of the cyclotron feature. Thus, rotational phase-resolved observations in principle can allow detailed study of the magnetic field geometry of the neutron star.

Much work as been performed to theoretically simulate the expected continuum spectrum from the shock as well as the shape and width of the cyclotron feature and allowing the characteristics of the neutron star magnetic field to be inferred \citep[see ][ and references therein]{schonherr2014}. However, the understanding of the CRSF shape and depth as a function of physical parameters --- e.g. magnetic field geometry, accretion rate, gas temperature --- is still not complete. A unified timing model is being constructed \citep{schonherr2014} but many physical effects, such as velocity and thermal gradients, plasma-magnetic field coupling need to be incorporated into the model before astrophysical parameters can be accurately inferred. In this work, we emphasize the discovery and observations of the cyclotron line in \rxj\, deferring the detailed modeling to a later date.  
 
\section{Observation and Analysis}
\label{sec:obs}
\rxj\ was first observed by \nustar\ on 2014 January 22 for a total exposure time of 27.8\,ks. The \nustar\ pointing was adjusted to avoid stray light from the bright X-ray source LMC X-1, but the alignment led to the source being incident on the gap between the CdZnTe detectors. The observation was repeated on 2014 January 24 with the pointing adjusted to center the source away from the detector gap. However, the data from both observations is usable as the response files created by the \nustar\ pipeline accurately correct for the effects of the detector gaps. The details of the observations are summarized in Table~\ref{tab:obs}. The observation numbers 80001002002 and 80001002004 are henceforth referred to as observations I and II, respectively. 

\begin{deluxetable}{lrrrr}
  \centering
  \tablecolumns{5}
  \tablecaption{\nustar\ observations of \rxj.\label{tab:obs}}
  \tablewidth{0pt}
  \tabletypesize{\footnotesize}
  \tablehead{
    \colhead{Obs ID}   &
    \colhead{Start}      &
    \colhead{End}   &
    \colhead{Exp}      &
    \colhead{Rate\tablenotemark{a}}\\
    \colhead{}  &
    \colhead{(UT)}                    &
    \colhead{(UT)}                    &
    \colhead{(ks)}                 &
    \colhead{cts/s}
  }
  \startdata
  80001002002 (I) & Jan 22 20:16 & Jan 23 11:36 & 27.7 & 9.5\\
  80001002004 (II)& Jan 24 23:56 & Jan 25 18:31 & 33.2 & 12\\
  \enddata
  \tablenotetext{a}{Average count-rate in each \nustar\ telescope. The count-rates are not corrected for photons lost in detector gaps.}
\end{deluxetable}

While the observations were taken during the long decline of the outburst, there are short-term variations in the lightcurve. \rxj\ was brighter during Observation II than during Observation I. We calculated the average background-subtracted count rate of \rxj\ from \swift-BAT orbital data over the \nustar\ observation periods. We find that the \swift-BAT count rates errors are 26\% and 37\% of the count rates respectively and while they are consistent with the \nustar\ observations, we cannot independently confirm the short-term brightening with \swift-BAT observations.

The preliminary processing and filtering of the \nustar\ event data was performed with the standard \nustar\ pipeline version 1.3.0 (\texttt{nupipeline}) and \texttt{HEASOFT} version 6.15. The source was clearly detected during each epoch across the entire 3--79\,keV. We used the \texttt{barycorr} tool to correct the photon arrival times to the barycenter of the solar system using the DE-200 ephemeris \citep{standish1992}. We extracted source events within a 40\,pixel (100\arcsec, compared to a half-power radius of $\approx$30\arcsec) radius around the centroid and suitable background regions were used for background estimation. Spectra were extracted using the \texttt{nuproducts} script. Using \texttt{grppha}, all photons below channel 35 (3\,keV) and above channel 1935 (79\,keV) were flagged as bad and all good photons were binned in energy to achieve a minimum of 30 photons per bin. All uncertainties quoted or plotted are 90\% confidence intervals ($\Delta\chi^{2}=2.706$) unless stated otherwise.

\section{Results}
\label{sec:results}

\subsection{Spectral Fitting}
The spectra extracted from both observations were fitted simultaneously with model spectra using \texttt{XSPEC} version 12.8.1i. The data from the two \nustar\ telescope were fitted linked by a cross-normalization factor. We extracted the background spectra from neighboring source-free regions on the same detector. The background rates were a factor of 10 fainter than the source spectra at 40\,keV and upto a factor of 1000 fainter at lower energies. We estimated the Galactic photoelectric absorption towards \rxj\ to be $1.8\times10^{21}\,\mathrm{cm^{-2}}$ from the Leiden/Argentine/Bonn (LAB) Survey of Galactic \ion{H}{1} \citep{kalberla2005} using the \texttt{HEASARC nh} tool. The spectral fits in the \nustar\ energy band were insensitive to the relatively small value of $N_\mathrm{H}$ and hence we froze $N_H$ to the Galactic value. There may be additional absorption from the Magellanic Cloud or intrinsic to the source, however we observe no difference in the spectral fitting with $N_\mathrm{H}=0\,\mathrm{cm^{-2}}$ and $N_\mathrm{H}=1.8\times10^{21}\,\mathrm{cm^{-2}}$ in the \texttt{XSPEC} absorption model \citep[\texttt{tbabs};][]{wilms2000}.

The continuum spectrum can be well-fit by three different power law models with a rollover: 1) a power law with a Fermi-Dirac cutoff \citep{tanaka1986}, 2) a thermally Comptonized continuum \citep[\texttt{nthcomp};][]{zycki1999} with a blackbody input photon spectrum, and 3) a power law with an exponential cutoff and an additional blackbody component. Table~\ref{tab:jan_spectra} (first section) shows the parameter values measured for the first model with a reduced $\chi^2$ of 1.5 for 1286 degrees of freedom (dof) for the first epoch and reduced $\chi^2_\mathrm{red}/\mathrm{dof} = 1.6/1424$ for the second epoch. These large $\chi^2_\mathrm{red}$ are typical of all continuum-only fits. 

\begin{figure}
\center
\includegraphics[clip=true,trim=0.2in 0in 0.2in 0in,width=0.48\textwidth]{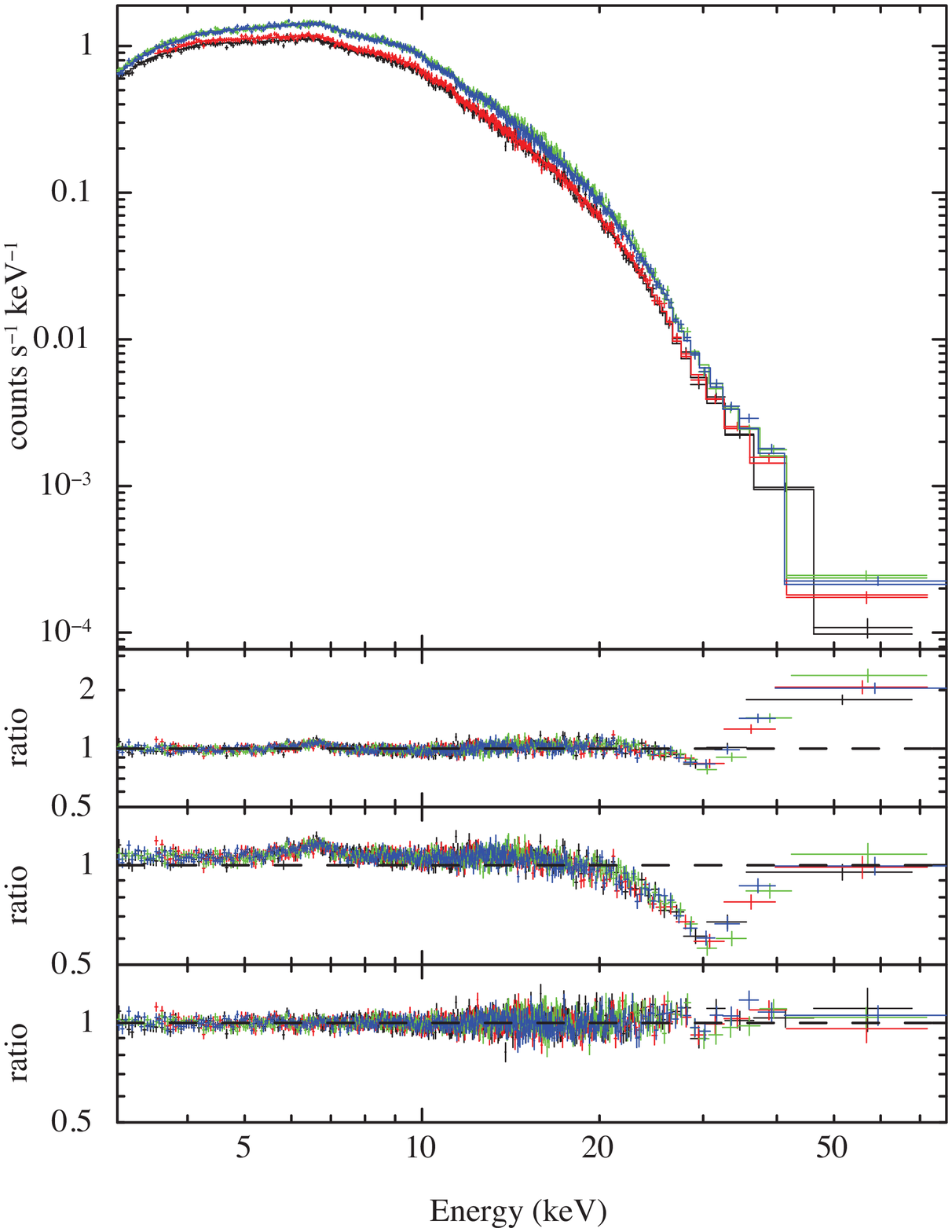}
\caption{\emph{First Panel (Top):} \nustar\ spectra from both observations. The colors correspond to the following data sets: black: Obs I \texttt{FPMA},  red: Obs I \texttt{FPMB}, green: Obs II \texttt{FPMA} and blue: Obs II \texttt{FPMB}. \emph{Second Panel:} Ratio of the continuum-only model --- \texttt{const*tbabs*cflux*(powerlaw*fdcut)} --- to the data. The Fe emission line at 6.5\,keV and the CRSF at $\approx$30\,keV are clearly visible. \emph{Third Panel:} Ratio of the full model --- \texttt{const*tbabs*cflux*(powerlaw*fdcut*gabs+gauss)} --- fit to the data plotted with the absorption and emission line strength set to zero. The continuum is accurately fit by the power law with an exponential cutoff. \emph{Fourth (Bottom) Panel:} Ratio of the full model --- \texttt{const*tbabs*cflux*(powerlaw*fdcut*gabs+gauss)} --- to the data.}
\label{fig:powerlaw_fdcut_gabs_gauss}
\end{figure}

Each continuum-only fit shows the presence of a deep, broad absorption feature centered at 30\,keV and an emission feature centered at 6.6\,keV (Figure~\ref{fig:powerlaw_fdcut_gabs_gauss}). The broad emission feature centered at 6.6\,keV is due to Fe K-shell emission and the deep broad absorption feature results from a cyclotron resonant scattering feature (CRSF) as we shall discuss in Section~\ref{sec:discussion}. Figure~\ref{fig:powerlaw_fdcut_gabs_gauss} (top panel) shows the best-fit model --- \texttt{const*tbabs*cflux*(powerlaw*fdcut*gabs+gauss)} --- fit to the data. The corresponding data-to-model ratio is shown in the bottom panel. 

The inclusion of the 30\,keV absorption feature (using the \texttt{XSPEC} model \texttt{gabs}) and the 6.6\,keV emission feature (\texttt{gauss}) drastically improves the $\chi_\mathrm{red}^2$ of the fitted models as shown in Table~\ref{tab:jan_spectra}. The best-fit achieved a $\chi^2_\mathrm{red}/\mathrm{dof} = 1.045/1280$ for the first epoch and reduced $\chi^2_\mathrm{red}/\mathrm{dof} = 1.006/1418$ for the second epoch. Based on the residual $\chi^2$ of all the fits from Table~\ref{tab:jan_spectra}, we adopt a \texttt{powerlaw*fdcut} continuum (`best-fit model') for all the following discussions unless otherwise specified. The \texttt{const*tbabs*cflux(cutoffPL*gabs+gauss+bbody)} model converges to a blackbody temperature $T\approx4$\,keV, which is significantly larger than usually observed for accreting neutron stars and hence we do not consider the third model to be physically plausible. The Fe emission line and CRSF parameters are relatively insensitive to the exact continuum model used, affirming the robust detection and measurement of these features.

\begin{deluxetable}{llcc}
\tablecolumns{4}
\tablecaption{Spectral Fits to \nustar\ observations.\label{tab:jan_spectra}}
\tablewidth{0pt}
\tabletypesize{\footnotesize}
\tablehead{
  \colhead{Component} &
  \colhead{Parameter}   &
  \multicolumn{2}{c}{Observation}\\
  \colhead{} &
  \colhead{} &
  \colhead{I}  &
  \colhead{II}
}
\startdata
\sidehead{\texttt{const*tbabs*cflux(powerlaw*fdcut)}}
\texttt{const}     & $C_\mathrm{FPMB}$\tablenotemark{a}                  &   $1.022_{-0.004}^{+0.004}$                  & $1.026_{-0.003}^{+0.003}$   \\
                    &  $\log_{10}(\mathrm{Flux})$\tablenotemark{b}      & $-8.917_{-0.002}^{+0.002}$                 &           $-8.879_{-0.001}^{+0.002}$ \\
\texttt{powerlaw} &  $\Gamma$                                           & $0.91_{-0.02}^{+0.03}$                    &          $0.76_{-0.02}^{+0.02}$       \\
\texttt{fdcut}    & $E_\mathrm{cutoff}$  (keV)                          & $12.9_{-0.6}^{+0.6}$                       &           $11.4_{-0.5}^{+0.5}$            \\
                  & $E_\mathrm{fold}$    (keV)                          & $6.2_{-0.1}^{+0.1}$                        &           $6.24_{-0.07}^{+0.07}$         \\
                  & $\chi^2/$dof                                       &   1.499/1286                                   &  1.589/1424   \\
\sidehead{\texttt{const*tbabs*cflux(powerlaw*fdcut*gabs+gauss)}}
\texttt{const}     & $C_\mathrm{FPMB}$\tablenotemark{a}               &   $1.022_{-0.004}^{+0.004}$                    &          $1.026_{-0.003}^{+0.003}$  \\
                  &  $\log_{10}(\mathrm{Flux})$\tablenotemark{b}      & $-8.910_{-0.002}^{+0.002}$                 &           $-8.874_{-0.002}^{+0.002}$ \\
\texttt{powerlaw} &  $\Gamma$                                        & $0.87_{-0.04}^{+0.04}$                        &          $0.74_{-0.03}^{+0.03}$       \\
\texttt{fdcut}    & $E_\mathrm{cutoff}$  (keV)                        & $10_{-2}^{+2}$                          &           $9_{-1}^{+1}$            \\
                  & $E_\mathrm{fold}$    (keV)                        & $7.9_{-0.3}^{+0.3}$                        &           $7.7_{-0.2}^{+0.3}$         \\
\texttt{gabs}     & $E_\mathrm{CRSF}$    (keV)                        & $31.3_{-0.7}^{+0.8}$                      &           $31.5_{-0.6}^{+0.7}$      \\
                  & $\sigma_\mathrm{CRSF}$ (keV)                      & $5.9_{-0.6}^{+0.7}$                        &           $5.8_{-0.5}^{+0.6}$     \\
                  & $\tau_\mathrm{CRSF}$\tablenotemark{d}             & $0.60_{-0.07}^{+0.08}$                      &           $0.57_{-0.07}^{+0.07}$   \\
\texttt{gauss}    & $E_\mathrm{Fe}$      (keV)                        & $6.58_{-0.05}^{+0.05}$                        &           $6.58_{-0.05}^{+0.05}$    \\
                  & $\sigma_\mathrm{Fe}$ (keV)                        & $0.38_{-0.02}^{+0.02}$                        &           $0.46_{-0.06}^{+0.07}$        \\
                  & norm\tablenotemark{c}                            & $1.1_{-0.2}^{+0.2}$                       &            $1.5_{-0.2}^{+0.2}$        \\
                  & $\chi^2/$dof                                     &   1.045/1280                                  & 1.006/1418    \\
\sidehead{\texttt{const*tbabs*cflux(nthcomp*gabs+gauss)}}
\texttt{const}     & $C_\mathrm{FPMB}$\tablenotemark{a}                & $1.022_{-0.004}^{+0.004}$                   &          $1.026_{-0.004}^{+0.004}$ \\
                   & $\log_{10}(\mathrm{Flux})$\tablenotemark{b}       & $-8.910_{-0.002}^{+0.002}$                 &          $-8.875_{-0.002}^{+0.002}$  \\
\texttt{nthcomp}  & $\Gamma$                                          & $1.46_{-0.01}^{+0.01}$                      &         $1.47_{-0.02}^{+0.01}$        \\
                  & $kT_\mathrm{e}$  (keV)                             & $5.1_{-0.2}^{+0.2}$                        &          $5.1_{-0.2}^{+0.2}$             \\
                  & $kT_\mathrm{bb}$    (keV)                          & $1.00_{-0.05}^{+0.04}$                      &          $0.84_{-0.05}^{+0.05}$  \\
\texttt{gabs}     & $E_\mathrm{CRSF}$    (keV)                         & $32_{-1}^{+1}$                      &           $32_{-1}^{+1}$            \\
                  & $\sigma_\mathrm{CRSF}$ (keV)                       & $8.6_{-0.9}^{+1.0}$                       &             $8.2_{-0.8}^{+1.2}$  \\
                  & $\tau_\mathrm{CRSF}$\tablenotemark{d}              & $0.8_{-0.1}^{+0.2}$                     &             $0.7_{-0.1}^{+0.2}$    \\
\texttt{gauss}    & $E_\mathrm{Fe}$      (keV)                         & $6.5_{-0.1}^{+0.1}$                   &            $6.62_{-0.05}^{+0.05}$  \\
                  & $\sigma_\mathrm{Fe}$ (keV)                         & $1.0_{-0.1}^{+0.2}$                     &             $0.51_{-0.08}^{+0.08}$           \\
                  & norm\tablenotemark{c}                             & $4.0_{-0.7}^{+1.0}$                      &            $5.5_{-0.3}^{+0.3}$             \\
                  &  $\chi^2/$dof                                     &   1.090/1280                                &        1.062/1418  \\
\sidehead{\texttt{const*tbabs*cflux(cutoffPL*gabs+gauss+bbody)}}
\texttt{const}     & $C_\mathrm{FPMB}$\tablenotemark{a}                &   $1.022_{-0.004}^{+0.004}$   &             $1.026_{-0.003}^{+0.003}$ \\
                   & $\log_{10}(\mathrm{Flux})$\tablenotemark{b}       & $-8.910_{-0.002}^{+0.002}$ &                $-8.874_{-0.002}^{+0.002}$  \\
\texttt{cutoffPL}   & $\Gamma$                                        & $0.6_{-0.1}^{+0.1}$        &               $0.7_{-0.1}^{+0.1}$      \\
                     & $E_{\mathrm{fold}}$ (keV)                        & $8.4_{-0.6}^{+0.8}$        &               $9.2_{-0.6}^{+0.7}$        \\
 \texttt{gabs}       & $E_\mathrm{CRSF}$ (keV)                         & $31_{-1}^{+1}$       &               $30.8_{-0.5}^{+0.6}$           \\
                     & $\sigma_\mathrm{CRSF}$ (keV)                    & $5.4_{-0.5}^{+0.5}$        &              $5.0_{-0.5}^{+0.5}$      \\
                  & $\tau_\mathrm{CRSF}$\tablenotemark{d}             & $1.2_{-0.3}^{+0.5}$                    &    $0.9_{-0.1}^{+0.1}$         \\
 \texttt{gauss}      & $E_\mathrm{Fe}$ (keV)                          & $6.59_{-0.05}^{+0.05}$      &              $6.57_{-0.05}^{+0.05}$     \\
                     & $\sigma_\mathrm{Fe}$ (keV)                     & $0.35_{-0.08}^{+0.09}$         &                $0.49_{-0.07}^{+0.08}$     \\
                     & norm ($10^{-2}$)\tablenotemark{c}             & $1.4_{-0.2}^{+0.2}$         &             $2.0_{-0.2}^{+0.2}$           \\
\texttt{bbody}      & $kT$ (keV)                                     & $4.5_{-0.3}^{+0.3}$         &              $4.0_{-0.1}^{+0.2}$         \\
                     & norm ($10^{-2}$)\tablenotemark{c}               & $9.1_{-0.4}^{+0.4}$         &             $11.7_{-0.8}^{+0.7}$ \\
                     & $\chi^2/$dof                                 &  1.046/1279                     & 1.006/1417 \\
\enddata
\tablecomments{The photo-electric absorption $N_\mathrm{H}$ was also frozen to the estimated Galactic value of $1.8\times10^{21}\,\mathrm{cm^{-2}}$.}
\tablenotetext{a}{The value of $C_\mathrm{FPMB}$ was tied across both epochs but allowed to vary for each fit. The value of \texttt{const} was frozen to unity for all FPMA data.}
\tablenotetext{b}{3--79\,keV flux in units of $\mathrm{erg\,cm^{-2}\,s^{-1}.}$ }
\tablenotetext{c}{Because the \texttt{cflux} model is used to compute the total flux, the normalization of the \texttt{gauss} model is dependent on the normalization of the powerlaw, which is frozen to $1\,\mathrm{phot\,keV^{-1}\,cm^{-2}\,s^{-1}}$ at 1\,keV. \texttt{norm} is specified in units of $10^{-2}\,\mathrm{phot\,cm^{-2}\,s^{-1}}$.}
\tablenotetext{d}{Optical depth of cyclotron absorption.}
\end{deluxetable}

The fitting of CRSF features is highly dependent on the model of the continuum. In fact \citet{muller2013} showed that the variation between $E_\mathrm{CRSF}$ and $L_X$ in 4U\,0115+634 (reported by \citet{nakajima2006}, \citet{tsygankov2007} and \citet{muller2010}) disappeared when the continuum model was improved. To explore the correlations between parameters, we stepped through the parameter values for the \texttt{const*tbabs*cflux*(powerlaw*fdcut*gabs+gauss)} model. Table~\ref{tab:continuum_variation} shows the typical variation of parameters as the power law index $\Gamma$ is frozen to values between 0.5 and 1.1. We observe that $\Gamma$ is tightly correlated with $E_\mathrm{cutoff}$. We observe that our measurement of the CRSF parameters is robust to variations in the continuum fitting. 

\begin{deluxetable*}{lllllll}
\tablecolumns{7}
\tablecaption{Effect of varying the continuum parameters\label{tab:continuum_variation}}
\tablewidth{0pt}
\tabletypesize{\footnotesize}
\tablehead{
  \colhead{} &
  \multicolumn{5}{c}{Fit Parameters} &
  \colhead{}\\
  \colhead{$\Gamma$} &
  \colhead{$E_\mathrm{cutoff}$} &
  \colhead{$E_\mathrm{fold}$} &
  \colhead{$E_\mathrm{CRSF}$} &
  \colhead{$\sigma_\mathrm{CRSF}$} &
  \colhead{$\tau_\mathrm{CRSF}$} &
  \colhead{$\chi^2_\mathrm{red}$\tablenotemark{a}}\\
  \colhead{} &
  \colhead{(keV)} &
  \colhead{(keV)} &
  \colhead{(keV)} &
  \colhead{(keV)} &
  \colhead{} &
  \colhead{}
}
\startdata
0.5&	0.01       &	 7.8    &    31.8&	5.1&	0.61&      	1.1078\\
0.6&	2.5	   &     8.1    &    32.2&	5.8&	0.67&      	1.0397\\
0.7&	7.2	   &     7.8    &    31.7&	5.7&	0.60&      	1.0094\\
$0.74_{-0.03}^{+0.03}$&	$8.9_{-1.3}^{+1.1}$&	$7.7_{-0.2}^{+0.2}$&	$31.5_{-0.6}^{+0.7}$&	$5.8_{-0.5}^{+0.6}$&	$0.57_{-0.06}^{+0.07}$&      	$1.0055$\\
0.8&	11.0	   &     7.5    &    31.5&	6.2&	0.56&      	1.0130\\
0.9&	16.4	   &     7.5    &    32.7&	8.9&	0.84&      	1.0496\\
1.0&	29.9	   &     5.0    &    31.9&	9.8&	1.74&      	1.0963\\
1.1&	31.5	   &     4.3    &    31.0&	8.1&	1.68&      	1.2767\\
\enddata
\tablenotetext{a}{The fit had 1419 degrees of freedom after freezing $\Gamma$.}
\tablecomments{The \nustar\ data for Observation II were fit with the \texttt{const*tbabs*cflux(powerlaw*fdcut*gabs+gauss)} model. The values of $\Gamma$ were set and frozen and the rest of the model parameters were fit. For Line 4, the value of $\Gamma$ was set to the best-fit value. The $\chi^2_\mathrm{red}$ value is reported after freezing $\Gamma$ for consistency. The 90\% confidence level error bars are reported.}
\end{deluxetable*}

\subsubsection{Cyclotron Harmonics}
Our data are very well fit by a single absorption line at $\approx$30\,keV. However, cyclotron resonance scattering may occur in multiple harmonics. We performed a check to confirm that the 30\,keV feature is the fundamental harmonic by adding a \texttt{gabs} component with the central energy at half that of the 30\,keV line. The width of the $\approx$15\, keV line was set to half the width of the second harmonic \citep[in the manner of ][]{furst2014_velax1}. The depth of the new component as well as all other components of the model were fitted. The fit nominally converged to the same parameter values as in Table~\ref{tab:jan_spectra} limiting the optical depth of the new line component to $\tau<0.008$ (90\% confidence) in individual observations and $\tau<0.004$ if the optical depth is tied between the two observations. Noting the factor of $\approx$50-80 difference between the strengths of the harmonics, we discard the possibility that the 30\,keV absorption feature is the second harmonic.

Similarly, we added a \texttt{gabs} model as the second harmonic of the 30\,keV absorption feature (with central energy and width double those of the fundamental) and re-fit the data. The optical depth of the absorption feature is very poorly constrained in individual observations: $\tau<2$ for Observation I and $\tau<1$ for Observation II. If the optical depth of the absorption is tied between the two observations, it is constrained to $\tau<1.4$. While a second harmonic line cannot be conclusively ruled out, it is not required for the spectrum to be fit. 

Table~\ref{tab:harmonic_fit} lists the best fit parameter values for both the above fits.  We note that all other parameters are consistent with the corresponding single absorption feature model from Table~\ref{tab:jan_spectra}.

\begin{deluxetable}{llcc}
\tablecolumns{4}
\tablecaption{Best-fit continuum model with CRSF and harmonic.\label{tab:harmonic_fit}}
\tablewidth{0pt}
\tabletypesize{\footnotesize}
\tablehead{
  \colhead{Component} &
  \colhead{Parameter}   &
  \multicolumn{2}{c}{Observation}\\
  \colhead{} &
  \colhead{} &
  \colhead{I}  &
  \colhead{II}
}
\startdata
\sidehead{\texttt{const*tbabs*cflux(powerlaw*fdcut*gabs*gabs+gauss)}} 
\texttt{const}     & $C_\mathrm{FPMB}$\tablenotemark{a}         &  $1.022_{-0.004}^{+0.004}$             &  $1.026_{-0.003}^{+0.003}$ \\
                   &  $\log_{10}(\mathrm{Flux})$\tablenotemark{b}      & $-8.910_{-0.002}^{+0.002}$      &    $-8.874_{-0.002}^{+0.002}$ \\
\texttt{powerlaw} &  $\Gamma$                       & $0.87_{-0.04}^{+0.04}$                        &          $0.74_{-0.03}^{+0.03}$       \\
\texttt{fdcut}    & $E_\mathrm{cutoff}$  (keV)       & $10_{-2}^{+2}$                          &           $9_{-1}^{+1}$            \\
                  & $E_\mathrm{fold}$    (keV)       & $7.9_{-0.3}^{+0.3}$                        &           $7.7_{-0.2}^{+0.2}$         \\
\texttt{gabs}     & $E_\mathrm{CRSF,f}$    (keV)       & $=E_\mathrm{CRSF,h}/2$                      &        $=E_\mathrm{CRSF,h}/2$      \\
                  & $\sigma_\mathrm{CRSF,f}$ (keV)     & $=\sigma_\mathrm{CRSF,h}/2$                 &        $=\sigma_\mathrm{CRSF,h}/2$      \\
                  & $\tau_\mathrm{CRSF,f}$\tablenotemark{d}               & $ <0.007$                 &             $<0.009$         \\
\texttt{gabs}     & $E_\mathrm{CRSF,h}$    (keV)       & $31.3_{-0.7}^{+0.8}$                       &           $31.5_{-0.6}^{+0.7}$      \\
                  & $\sigma_\mathrm{CRSF,h}$ (keV)     & $5.9_{-0.6}^{+0.7}$                           &           $5.8_{-0.5}^{+0.6}$     \\
                  & $\tau_\mathrm{CRSF,h}$\tablenotemark{d}    & $0.60_{-0.07}^{+0.08}$                &           $0.57_{-0.06}^{+0.07}$   \\
\texttt{gauss}    & $E_\mathrm{Fe}$      (keV)       & $6.58_{-0.05}^{+0.05}$                      &           $6.58_{-0.05}^{+0.05}$    \\
                  & $\sigma_\mathrm{Fe}$ (keV)       & $0.39_{-0.08}^{+0.09}$                        &           $0.46_{-0.06}^{+0.07}$        \\
                  & norm\tablenotemark{c}                & $1.1_{-0.2}^{+0.2}$                        &            $1.5_{-0.2}^{+0.2}$        \\
                  & $\chi^2/$dof                   &   1.046/1279        &  1.009/1417  \\

\sidehead{\texttt{const*tbabs*cflux(powerlaw*fdcut*gabs*gabs+gauss)}} 
\texttt{const}     & $C_\mathrm{FPMB}$\tablenotemark{a}         &  $1.022_{-0.004}^{+0.004}$                              & $1.026_{-0.003}^{+0.003}$   \\
                  &  $\log_{10}(\mathrm{Flux})$\tablenotemark{b}      & $-8.911_{-0.002}^{+0.002}$                &           $-8.874_{-0.002}^{+0.001}$ \\
\texttt{powerlaw} &  $\Gamma$                       & $0.84_{-0.05}^{+0.05}$                        &          $0.74_{-0.04}^{+0.04}$       \\
\texttt{fdcut}    & $E_\mathrm{cutoff}$  (keV)       & $8_{-2}^{+3}$                          &           $9_{-2}^{+2}$            \\
                  & $E_\mathrm{fold}$    (keV)       & $8.4_{-0.6}^{+0.4}$                        &           $7.6_{-0.2}^{+0.6}$         \\
\texttt{gabs}     & $E_\mathrm{CRSF,f}$    (keV)       & $31.6_{-1}^{+1}$                       &           $31.5_{-0.6}^{+0.8}$      \\
                  & $\sigma_\mathrm{CRSF,f}$ (keV)     & $6.2_{-0.7}^{+0.7}$                           &           $5.7_{-0.5}^{+0.8}$     \\
                  & $\tau_\mathrm{CRSF,f}$\tablenotemark{d}  & $0.7_{-0.1}^{+0.1}$                      &           $0.56_{-0.06}^{+0.15}$   \\
\texttt{gabs}     & $E_\mathrm{CRSF,h}$    (keV)       & $=2\,E_\mathrm{CRSF,f}$                       &           $=2\,E_\mathrm{CRSF,f}$          \\
                  & $\sigma_\mathrm{CRSF,h}$ (keV)     & $=2\,\sigma_\mathrm{CRSF,f}$                           &  $=2\,\sigma_\mathrm{CRSF,f}$    \\
                  & $\tau_\mathrm{CRSF,h}$\tablenotemark{d} & $ <2.0$                                     &           $<1.0$         \\
\texttt{gauss}    & $E_\mathrm{Fe}$      (keV)       & $6.58_{-0.05}^{+0.05}$                      &           $6.58_{-0.06}^{+0.05}$    \\
                  & $\sigma_\mathrm{Fe}$ (keV)       & $0.38_{-0.07}^{+0.09}$                        &           $0.46_{-0.06}^{+0.05}$        \\
                  & norm\tablenotemark{c}           & $1.0_{-0.2}^{+0.2}$                        &            $1.5_{-0.2}^{+0.1}$        \\
                  & $\chi^2/$dof                   &   1.045/1279                               &   1.007/1417 \\
\enddata
\tablecomments{The subscripts `f' and `h' refer to the fundamental line and the harmonic, respectively. The photo-electric absorption $N_\mathrm{H}$ was also frozen to the estimated Galactic value of $1.8\times10^{21}\,\mathrm{cm^{-2}}$.}
\tablenotetext{a}{The value of $C_\mathrm{FPMB}$ was tied across both epochs but allowed to vary for each fit. The value of \texttt{const} was frozen to unity for all FPMA data.}
\tablenotetext{b}{3--79\,keV flux in units of $\mathrm{erg\,cm^{-2}\,s^{-1}.}$ }
\tablenotetext{c}{Because the \texttt{cflux} model is used to compute the total flux, the normalization of the \texttt{gauss} model is dependent on the normalization of the powerlaw, which is frozen to $1\,\mathrm{phot\,keV^{-1}\,cm^{-2}\,s^{-1}}$ at 1\,keV. \texttt{norm} is specified in units of $10^{-2}\,\mathrm{phot\,cm^{-2}\,s^{-1}}$.}
\tablenotetext{d}{Optical depth of cyclotron absorption.}
\end{deluxetable}

\subsubsection{Cyclotron Absorption Profile}
To explore the shape of the absorption profile, we fit the spectra at both epochs replacing the Gaussian absorption line (\texttt{gabs}) with a Lorentzian absorption line specified by the XSPEC model \texttt{cyclabs}. Based on the previous discussion, we set the depth of the second harmonic to zero. The goodness of fit was nominally equivalent to that of the best-fit \texttt{const*tbabs*cflux(powerlaw*fdcut*gabs+gauss)} model. Table~\ref{tab:lorentz_fit} specifies the best-fit parameters and their errors. We note that $E_\mathrm{CRSF}$ systematically converged to a lower value than with the Gaussian profile, as previously noted for other sources by \citet{mihara1995} and many subsequent authors \citep[see ][ and references therein]{hemphill2013}. 

\begin{deluxetable}{llcc}
\tablecolumns{4}
\tablecaption{Best-fit continuum model with Lorentzian profile absorption\label{tab:lorentz_fit}}
\tablewidth{0pt}
\tabletypesize{\footnotesize}
\tablehead{
  \colhead{Component} &
  \colhead{Parameter}   &
  \multicolumn{2}{c}{Observation}\\
  \colhead{} &
  \colhead{} &
  \colhead{I}  &
  \colhead{II}
}
\startdata
\sidehead{\texttt{const*tbabs*cflux(powerlaw*fdcut*cyclabs+gauss)}} 
\texttt{const}     & $C_\mathrm{FPMB}$\tablenotemark{a}           &  $1.022_{-0.004}^{+0.004}$                              &   $1.026_{-0.003}^{+0.003}$ \\
                  &  $\log_{10}(\mathrm{Flux})$\tablenotemark{b}  & $-8.910_{-0.002}^{+0.002}$                &           $-8.874_{-0.002}^{+0.002}$ \\
\texttt{powerlaw} &  $\Gamma$                       & $0.88_{-0.04}^{+0.04}$                        &          $0.74_{-0.03}^{+0.03}$       \\
\texttt{fdcut}    & $E_\mathrm{cutoff}$  (keV)       & $10_{-2}^{+2}$                          &           $8_{-2}^{+1}$            \\
                  & $E_\mathrm{fold}$    (keV)       & $8.6_{-0.5}^{+0.6}$                        &           $8.3_{-0.4}^{+0.5}$         \\
\texttt{cyclabs}  & $E_\mathrm{CRSF}$    (keV)       & $28.6_{-0.5}^{+0.5}$                       &           $29.0_{-0.4}^{+0.4}$      \\
                  & $\sigma_\mathrm{CRSF}$ (keV)     & $9_{-1}^{+2}$                           &           $8.9_{-1}^{+2}$     \\
                  & $\tau_\mathrm{CRSF}$             & $0.7_{-0.1}^{+0.1}$                        &           $0.7_{-0.09}^{+0.10}$         \\
\texttt{gauss}    & $E_\mathrm{Fe}$      (keV)       & $6.58_{-0.05}^{+0.05}$                      &           $6.58_{-0.05}^{+0.05}$    \\
                  & $\sigma_\mathrm{Fe}$ (keV)       & $0.39_{-0.08}^{+0.10}$                        &           $0.47_{-0.06}^{+0.07}$        \\
                  & norm\tablenotemark{c}                & $1.11_{-0.2}^{+0.2}$                        &            $1.50_{-0.3}^{+0.3}$        \\
                  & $\chi^2/$dof                   &   1.049/1280                              &   1.008/1418 \\
\enddata
\tablecomments{The photo-electric absorption $N_\mathrm{H}$ was also frozen to the estimated Galactic value of $1.8\times10^{21}\,\mathrm{cm^{-2}}$.}
\tablenotetext{a}{The value of $C_\mathrm{FPMB}$ was tied across both epochs but allowed to vary for each fit. The value of \texttt{const} was frozen to unity for all FPMA data.}
\tablenotetext{b}{3--79\,keV flux in units of $\mathrm{erg\,cm^{-2}\,s^{-1}.}$ }
\tablenotetext{c}{Because the \texttt{cflux} model is used to compute the total flux, the normalization of the \texttt{gauss} model is dependent on the normalization of the powerlaw, which is frozen to $1\,\mathrm{phot\,keV^{-1}\,cm^{-2}\,s^{-1}}$ at 1\,keV. \texttt{norm} is specified in units of $10^{-2}\,\mathrm{phot\,cm^{-2}\,s^{-1}}$.}
\tablenotetext{d}{Optical depth of cyclotron absorption.}
\end{deluxetable}

 \subsection{Timing Measurements}
Using the rotational period estimate and instantaneous rotational period derivative (fixed) from \citet{kuehnel2014ATel}, we calculated the best-fit rotation period for both epochs using the \texttt{xronos} tool \texttt{efsearch} and the barycenter corrected data. The measured best-fit periods, 8.032375(5)\,s and 8.032932(5)\,s for observations I and II, respectively, are consistent with the ephemeris measured from \fermi. As each observation spanned $\approx$60\,ks, corresponding to 3\,\% of the orbital period, the expected pulse smearing due to the change in pulse period is negligible and was therefore ignored.  

\begin{figure*}
 \center
 \includegraphics[clip=true,trim=0.25in 0in 0.2in 0in,width=0.49\textwidth]{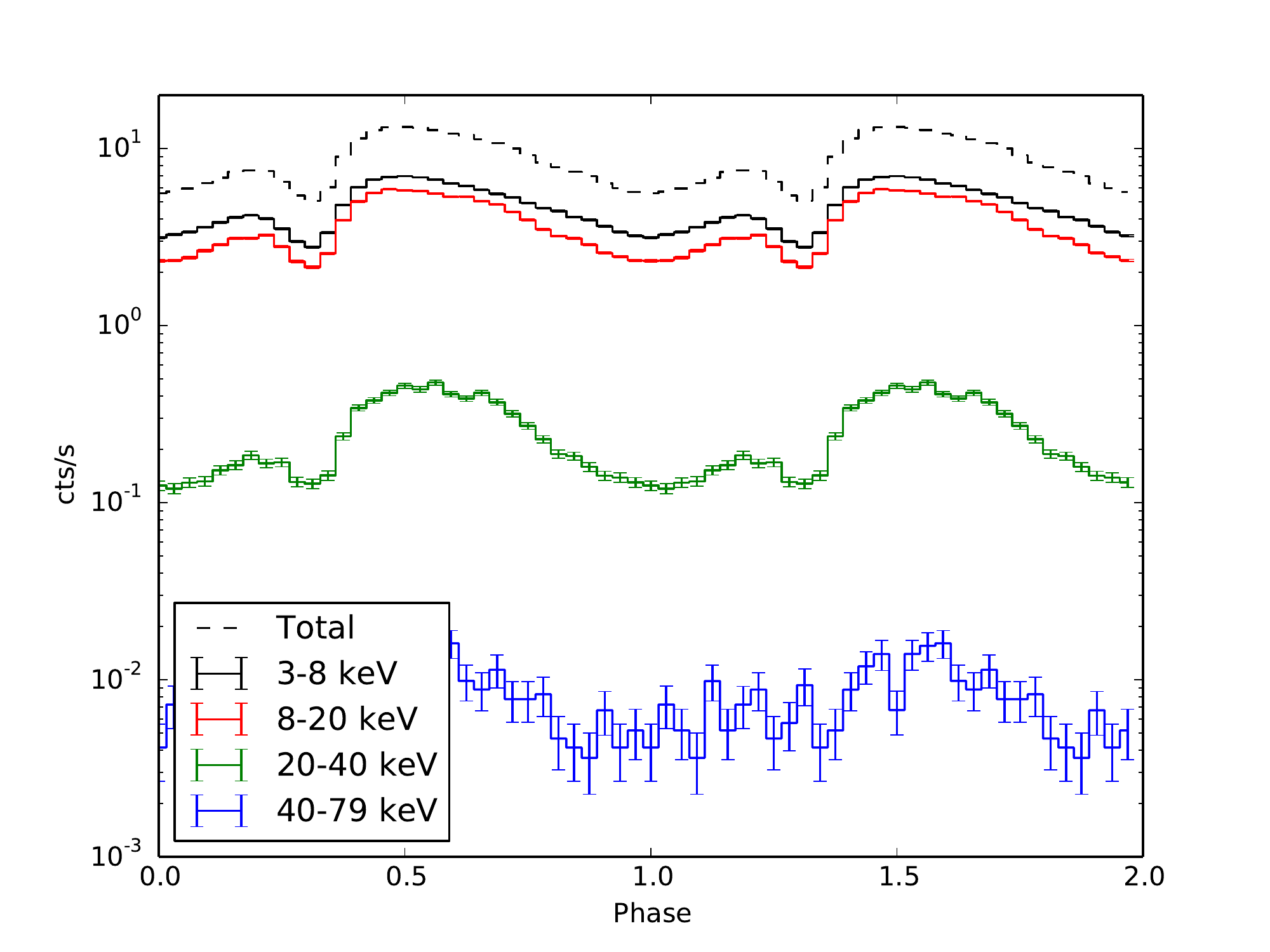}
 \includegraphics[clip=true,trim=0.25in 0in 0.2in 0in,width=0.49\textwidth]{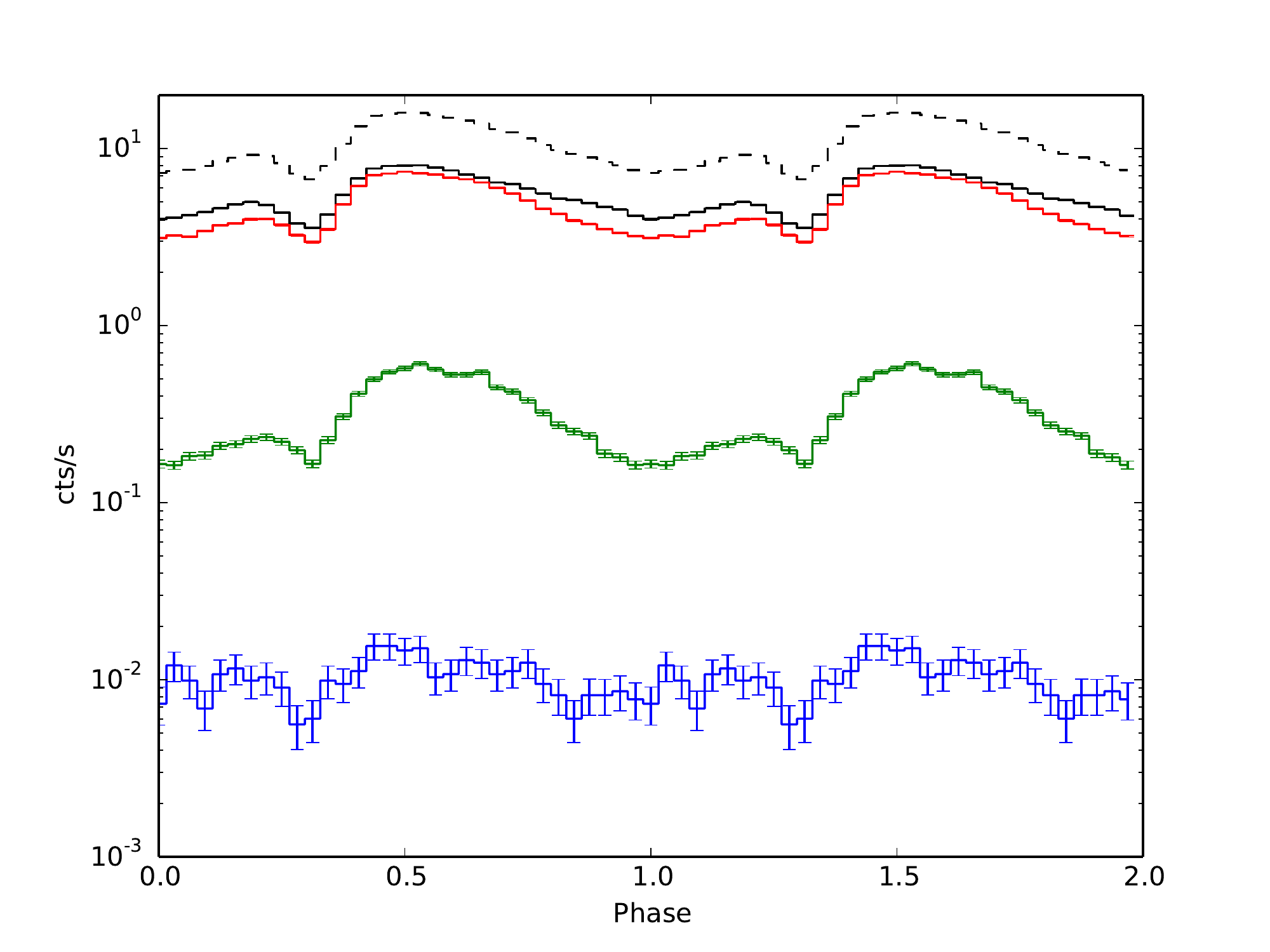}
 \caption{Pulse profile as a function of rotational phase for both Observation I (\emph{Left Panel}) and II (\emph{Right Panel}). Two pulses are shown for clarity. The dotted line shows the total pulse profile from 3--79\,keV. The black, red, green and blue solid lines denote the 3--8\,keV, 8--20\,keV, 20--40\,keV and 40--79\,keV pulse profiles, respectively. All error bars are 1-$\sigma$; for large count rates, error bars are too small to be visible. (This figure is available in color in the electronic version.)}
 \label{fig:pulse_color}
 \end{figure*}

The lightcurve for each observation was folded into 32 phase bins using the best-fit rotational period and instantaneous period derivative. To enable comparison between the two folded profiles, we chose the starting epochs to be MJD 56679.856670 and MJD 56682.010896, respectively, such that the peak 3--79\,keV flux is assigned a rotational phase of 0.5. We looked for variations in the pulse shape as a function of energy by folding the lightcurve in energy bins from 3--8\,keV, 8--20\,keV, 20--40\,keV and 40--79\,keV. Figure~\ref{fig:pulse_color} shows the pulse shapes in different energy bins for each observation. 

The 3--8\,keV, 8--20\,keV, and 20--40\,keV pulse profiles show a smooth, rapidly rising and slowly falling main pulse at a phase of 0.5 and a secondary pulse with a flipped symmetry (i.e. slowy rising and fast falling) at a phase of 0.2. The count rate in the 40--79\,keV profile is too low to distinguish asymmetry in the main pulse and locate a secondary pulse. We note that the dip between the primary peak and the secondary peak at a rotational phase of $\approx$0.28 is significantly sharper in Observation II than in Observation I (especially in the 20--40\,keV pulse profile). A small bump is detected in the 20--40\,keV pulse profile at a phase of 0.66 in both the observations. 

\subsubsection{Search for Other Periodic Features}
In order to search for any further possible periodic oscillations in the lightcurve, we calculated the power spectra of the two observations between 0.003--100\,Hz with a resolution of 0.0153\,Hz using the \texttt{xronos} tool \texttt{powspec}. Apart from sharp peaks at the rotational frequency ($\approx$0.1245\,Hz) and its harmonics, no other features were observed.

\subsection{Phase-Resolved Measurements}
In order to measure the spectral variations as a function of rotational phase, we created \emph{good-time-intervals} (\texttt{gti}s) based on the folding epoch and rotational period described above and extracted the spectra for 10 equal rotational phase bins using \texttt{nuproducts}. Photons in spectral channels were not binned and Cash statistics \citep[\texttt{cstat} in \texttt{XSPEC}; ][]{cash1979} were utilized to fit the model to the data. 

We used the model that best-fit the the average data, \texttt{const} \texttt{*} \texttt{tbabs} \texttt{*} \texttt{cflux} \texttt{(powerlaw} \texttt{*} \texttt{fdcut} \texttt{*} \texttt{gabs} \texttt{+gauss)}, to fit the phase-resolved spectra. Twenty spectra (10 phase bins for each of FPMA and FPMB) of each observation were fit simultaneously. As in the fitting of the average spectra, FPMB data were scaled by a cross normalization factor ($=$1.053) which was tied across all phases. After preliminary fits, it was observed that the \texttt{fdcut} parameters ($E_\mathrm{cutoff}$ and $E_\mathrm{fold}$) and the Fe emission line (\texttt{gauss}) parameters ($E_\mathrm{Fe}$, $\sigma_\mathrm{Fe}$, and normalization) did not vary at a statistically significantly level over the rotational phase. While these parameters have been observed to vary with rotational phase in other X-ray binaries (for example: Her X-1, \citealt{furst2013_herx1}, Cen X-3, \citealt{suchy2008} and GX\,301$-$2, \citealt{islam2014}), there was no statistical change in the goodness-of-fit when these parameters were tied across rotational phases. Freezing these parameters does not change the best-fit CRSF parameter values. As demonstrated above, any residual variation in $E_\mathrm{cutoff}$ is degenerate with the variation of $\Gamma$ and does not affect the CRSF parameters. 

The CRSF parameters for phase bins 0.0--0.1, 0.8--0.9 and 0.9--1.0 (corresponding to the low count-rate in Figure~\ref{fig:pulse_color}) were not well constrained. Hence we tied the parameters for these phase bins together to increase the signal-to-noise ratio in that bin and improve parameter constraints. The CRSF parameters for phase bin 0.7--0.8 in epoch 2 did not converge to physical values. However, the similarity of all CRSF parameters in every other phase bin for both the epochs suggests that this is not a physical disappearance of the cyclotron line. 

\begin{figure*}
\center
\includegraphics[clip=true,trim=0.2in 0in 0.2in 0in,width=0.48\textwidth]{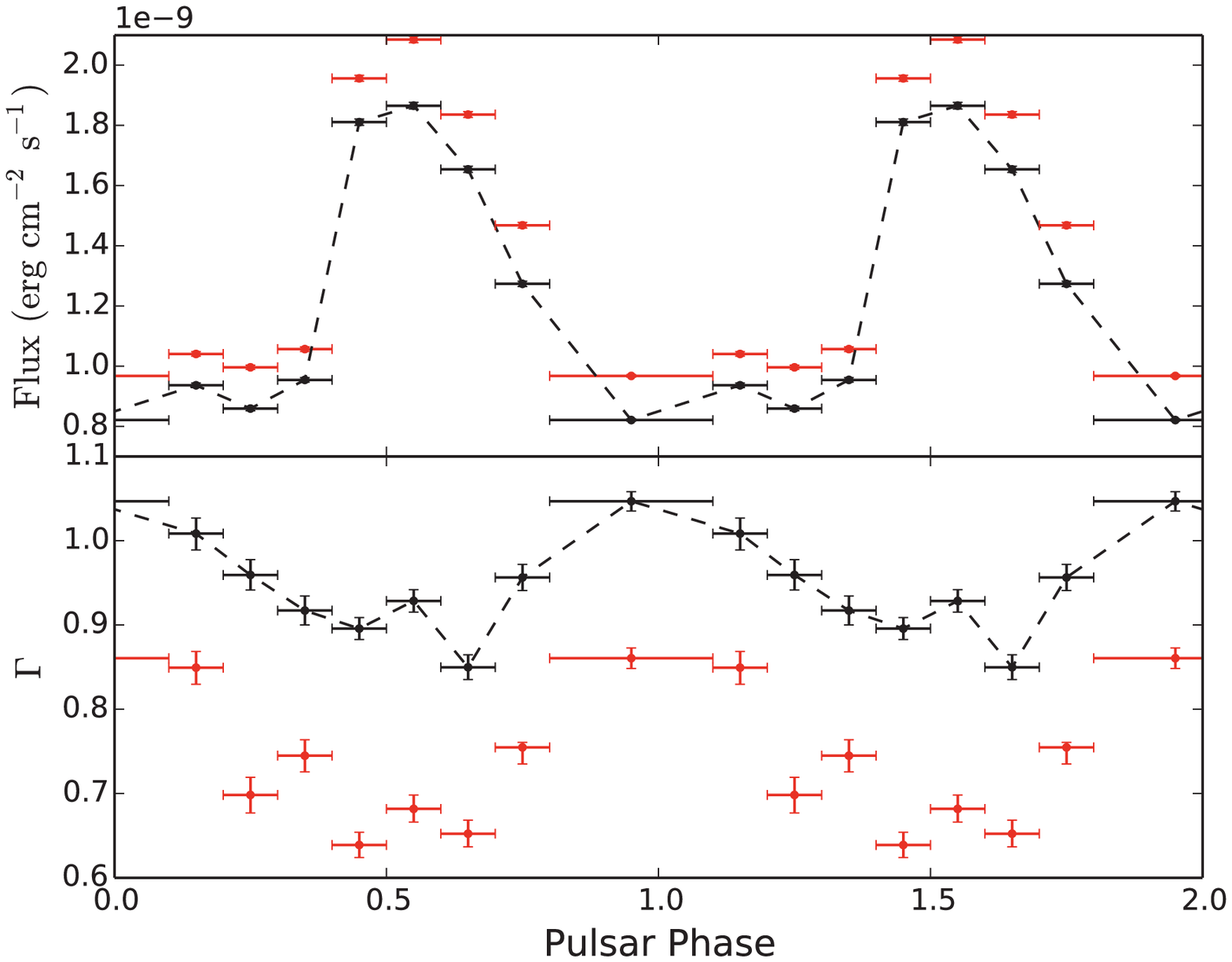}
\includegraphics[clip=true,trim=0.2in 0in 0.2in 0in,width=0.48\textwidth]{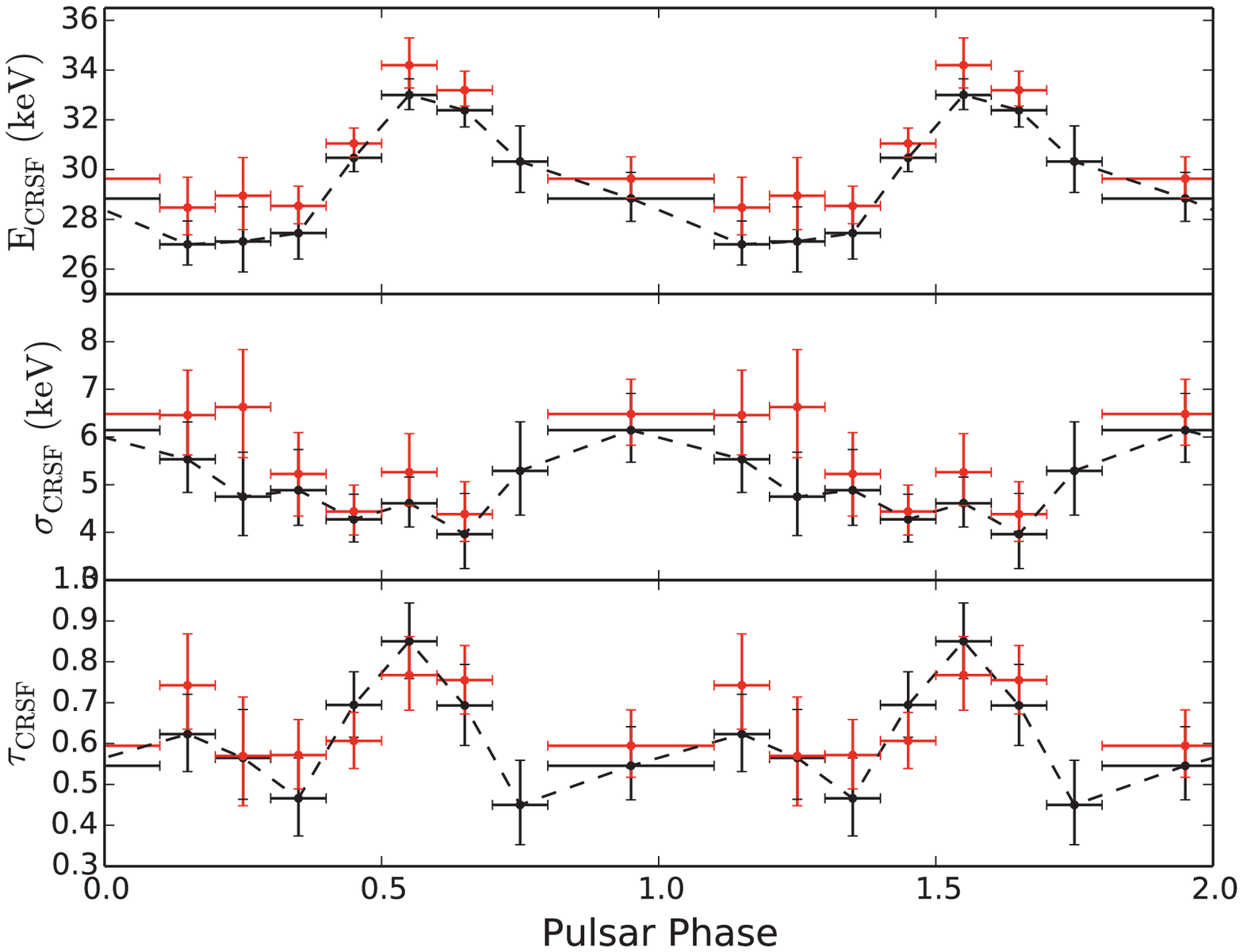}
\caption{Spectral parameter values as a function of phase. In each plot, black points are parameter values for Observation I and red points are for Observation II. The black points are connected with dashed black lines and the rotational cycle is plotted twice for clarity. 1-$\sigma$ errors are marked for each parameter. \emph{Left Panel Top:} Variation of 3--79\,keV flux as a function of rotational phase. The error bars in flux are typically $0.4\%$, too small to be seen on this plot. \emph{Left Panel Bottom:} Variation of power law index $\Gamma$. \emph{Right Panel: Top, Middle and Bottom}: Variation of CRSF central energy, width and optical depth, respectively. The central energy peaks in phase with the luminosity, similar to observations in most other pulsars. (This figure is available in color in the electronic version.)}
\label{fig:phase_resolved_spectra}
\end{figure*}

\begin{deluxetable}{llcc}
\centering
\tablecolumns{4}
\tablecaption{Parameters independent of rotational phase.\label{tab:phase_resolved_spectra}}
\tablewidth{0pt}
\tabletypesize{\footnotesize}
\tablehead{
  \colhead{Component} &
  \colhead{Parameter} &
  \multicolumn{2}{c}{Observation}\\
  \colhead{} &
  \colhead{} &
  \colhead{I} &
  \colhead{II} 
}
\startdata
\texttt{const}    & constant                                         & $1.066_{-0.003}^{+0.003}$ & $1.054_{-0.001}^{+0.004}$ \\
\texttt{fdcut}    & $E_\mathrm{cutoff}$  (keV)                        & $17.41_{-0.61}^{+0.44}$   & $15.54_{-0.42}^{+0.23}$  \\
                  & $E_\mathrm{fold}$    (keV)                        & $6.17_{-0.17}^{+0.15}$    & $5.94_{-0.06}^{+0.25}$  \\ 
\texttt{gauss}    & $E_\mathrm{Fe}$      (keV)                        & $6.573_{-0.031}^{+0.027}$ & $6.510_{-0.041}^{+0.045}$ \\ 
                  & $\sigma_\mathrm{Fe}$ (keV)                        & $0.479_{-0.052}^{+0.045}$ & $0.615_{-0.066}^{+0.024}$  \\ 
                  & norm ($10^{-2}$)                                  & $1.33_{-0.12}^{+0.10}$    & $2.37_{-0.36}^{+0.04}$    \\ 
\enddata
\tablecomments{The \texttt{const} parameter was frozen to unity for FPMA and left free for all observations with FPMB to allow for cross calibration errors. The photo-electric absorption $N_\mathrm{H}$ was frozen to the estimated Galactic value of $1.8\times10^{21}\,\mathrm{cm^{-2}}$.}
\end{deluxetable}

The fitting involved 46 free parameters over 20 spectra for each observation. After the fitting, we ran $\approx$830,000 Markov Chain Monte Carlo simulations for each observation. The errors on each free parameter were calculated by marginalizing over this data set. Table~\ref{tab:phase_resolved_spectra} lists the values for all parameters that were found to be independent of the rotational phase and Figure~\ref{fig:phase_resolved_spectra} shows the variations of the other parameters with rotational phase. Due to low photon statistics, we did not search for the cyclotron line harmonics at $\approx15$\,keV or $\approx60$\,keV, which may show up at specific rotational phases. However, a visual inspection of the fitted residuals shows no indication of the presence of such a feature at any phase.

\section{Discussion and Conclusions}
\label{sec:discussion}
We have presented spectral and timing analysis of two observations of Be/X-ray binary \rxj\ taken during its outburst in January 2014. Through the detection of a deep cyclotron resonant scattering feature at $\approx$30\,keV, we measure the magnetic field at the neutron star surface to be $2\times10^{12}$\,G. We robustly detect the CRSF in phase-averaged spectra  as well as in almost all phase-resolved spectra. We also detect an Fe K-shell emission line at 6.5\,keV in each observation. We now briefly discuss these observations in comparison with other Be/X-ray binary systems. 

\subsection{Broad Fe Emission Line}
We detected an Fe emission line with an intensity of $9_{-2}^{+5} \times 10^{-11}\,\mathrm{erg\,cm^{-2}\,s^{-1}}$ and $17_{-6}^{+4} \times 10^{-11}\,\mathrm{erg\,cm^{-2}\,s^{-1}}$ during observations I and II respectively. This corresponds to a luminosity of $2.7 \times 10^{37}\,\mathrm{erg\,s^{-1}}$ and $5.1 \times 10^{37}\,\mathrm{erg\,s^{-1}}$ at a nominal distance to the Large Magellanic Cloud of 50\,kpc \citep{inno2013}. We measure the width of the Fe emission line to be $0.38_{-0.02}^{+0.02}$\,keV for the first observation and  $0.46_{-0.07}^{+0.06}$\,keV for the second observation. This broadening is larger than typical for most neutron star binaries but not unphysical. In Her X-1, with \nustar\ and \emph{Suzaku} data, \citet{furst2013_herx1} resolved the Fe emission line into broad ($\sigma\approx$0.82\,keV) and narrow ($\sigma\approx$0.25\,keV) components. KS\,1947+300 has an Fe line with a width in the range 0.25--0.31\,keV \citep{furst2014_ks1947}. 

If a significant fraction of the broadening is caused by rotational broadening, the 0.45\,keV spread requires that the Fe emission line originates at distance of $\sim450$\,km from the neutron star, where the orbital velocity-scale is $\sim2 \times 10^{9}\,\mathrm{cm\,s^{-1}}$. However, it is likely that the Fe emission line in \rxj\ is broadened through a combination of rotational or thermal broadening and is a combination of multiple ionization states. We do not detect any detailed features in the line profile that may indicate multiple ionization states. 

\subsection{Pulse Profile}
We observe a sharply peaked pulse profile with a single dominant peak and a weak secondary peak, similar to the pulse profiles of GX\,301-2 \citep{suchy2012,islam2014} and Her\,X-1 \citep{furst2013_herx1}, and in contrast to the broad pulse profiles observed in KS\,1947+300 \citep{furst2014_ks1947}. The pulse profile does not evolve significantly between the two observing epochs, other than in total luminosity. 

The radiation from an accretion column in a dipolar magnetic field is generally expected to be symmetric around the magnetic axis. A decomposition of the pulse shape into two `similar', symmetric, and non-negative components, each originating from one polar cap, has been applied to various X-ray pulsar profiles \citep[see ][and references therein]{caballero2011,sasaki2012}. Our observation of an asymmetric pulse shape indicates that the magnetic dipole is likely offset from the rotation axis of the pulsar, hence the difference in longitude of each polar region is not $\pi$ radians. 

\subsection{Cyclotron Resonant Scattering Feature}
We observed a significant CRSF with a mean central energy of $31.40$\,keV, corresponding to a magnetic field strength of $2\times10^{12}$\,G. Using the formalism in \citep{becker2012}, we estimate the critical luminosity $L_\mathrm{crit}=1.5\times10^{37}\,B_\mathrm{12}^{16/15}\,\mathrm{erg\,s^{-1}}$ to be $3\times10^{37}\,\mathrm{erg\,s^{-1}}$. Assuming a distance of 50\,kpc, the luminosity of \rxj\ is $3.5\times10^{38}\,\mathrm{erg\,s^{-1}}$, far exceeding $L_\mathrm{crit}$.

At the observed luminosity, the accretion shock is expected to be radiation dominated, hence $E_\mathrm{CRSF}$ should decrease with increasing luminosity. Between Observations I and II, the luminosity increases by $\approx$9\% and we observe a $<0.6$\% increase in $E_\mathrm{CRSF}$. However, two aspects need to be noted: (1) the increase in $E_\mathrm{CRSF}$ has less than 1-$\sigma$ significance and (2) as reported by \citet{tsygankov2006} (in their Figure 4) the $E_\mathrm{CRSF}$ measurements in V0332+53 showed a scatter of upto 0.5\,keV for the same intrinsic luminosity. Similar to KS\,1947+300 \citep{furst2014_ks1947}, further observations over a larger luminosity range are required to determine if the expected correlation between $E_\mathrm{CRSF}$ and $L_X$ holds.
   
\subsubsection{Variation with Rotational Phase}
We observe a clear variation in the cyclotron absorption line as a function of rotation phase (Figure~\ref{fig:phase_resolved_spectra}). We observe a significant increase in $E_\mathrm{CRSF}$ as a function of luminosity. The $\chi^2$ value of the variation is 62.9 for observation I with 7 dof and 46.2 for observation II with 6 dof. The $E_\mathrm{CRSF}$ profile is similar in phase and shape to the count rate profile. We detect a very small lag of $\Delta\phi\approx0.1$ in the peak of the $E_\mathrm{CRSF}$ profile as compared to the count rate. Similar small or no lags have been observed in GX\,301$-$2 \citep[$\Delta\phi\approx0.2$; ][]{suchy2012}, Cen X-3 \citep[$\Delta\phi\approx0.1$; ][]{suchy2008} and Her X-1 \citep[$\Delta\phi\approx0$; ][]{furst2013_herx1}. 

We observe no variation in $\sigma_\mathrm{CRSF}$ in observation I with $\chi^2$/dof $=8.1/7$, but a marginal evidence for variation during observation II with $\chi^2$/dof $=12.3/6$. Similar to other systems, $\sigma_\mathrm{CRSF}$ has a slight peak at the minimum of the count rate. 

Similarly, we observe variation in the optical depth of observation, $\tau_\mathrm{CRSF}$, during observation I with $\chi^2$/dof $=14.7/7$ but not during observation II with $\chi^2$/dof $=6.4/6$. $\tau_\mathrm{CRSF}$ peaks at the maximum of $E_\mathrm{CRSF}$ and the count rate. 

\subsubsection{Cyclotron Line Shape}
The spectral profile of the CRSF depends on the detailed geometry of the magnetic field, the emission region and the path taken by photons through the plasma. In simulations, \citet{schonherr2007} showed that emission wings around CRSFs are supressed when the continuum spectrum has a low value of $E_\mathrm{fold}$ ($\approx5$\,keV) as observed in these observations. We do not find any evidence for emission wings around the CRSF feature. The data are equally well described by  a Gaussian or a Lorentzian absorption profile, in agreement with previous CRSF detections \citep[for example,][]{furst2013_herx1}. As the shock is radiation-dominated, most of the photons are expected to escape from the fan-beam (i.e. through the sides of the cylindrical accreting column).

With physically realistic theoretical models, these phase resolved CRSF observations of \rxj\ and other systems may be used in the near future as an excellent tool for the study of accretion physics, neutron star magnetic field geometry and neutron star crusts. The detailed physical modelling of this system is thus deferred to a later date.

\acknowledgements
We thank the anonymous referee for detailed suggestions and comments. This work was supported under NASA Contract No. NNG08FD60C, and made use of data from the {\it NuSTAR} mission, a project led by the California Institute of Technology, managed by the Jet Propulsion Laboratory, and funded by the National Aeronautics and Space Administration. We thank the {\it NuSTAR} Operations, Software and Calibration teams for support with the execution and analysis of these observations.  This research has made use of the {\it NuSTAR} Data Analysis Software (NuSTARDAS) jointly developed by the ASI Science Data Center (ASDC, Italy) and the California Institute of Technology (USA).

\newpage


\bibliographystyle{apj}
\bibliography{paper}

\end{document}